# Using Apple Machine Learning Algorithms to Detect and Subclassify Non-Small Cell Lung Cancer

Andrew A. Borkowski, MD*[1,2], Catherine P. Wilson, MT[1], Steven A. Borkowski[1], Lauren A. Deland, RN[1], Stephen M. Mastorides, MD[1,2]

[1]Pathology and Laboratory Medicine Service, James A. Haley VA Hospital, Tampa, Florida, USA.
[2]Department of Pathology and Cell Biology, University of South Florida, Morsani College of Medicine, Tampa, Florida, USA.

*aborkows@health.usf.edu
Pathology and Laboratory Medicine Service (673/113)
James A Haley VA Hospital
13000 Bruce B. Downs Blvd
Tampa, FL 33612
Tel: 813-972-7525

**ABSTRACT**

Lung cancer continues to be a major healthcare challenge with high morbidity and mortality rates among both men and women worldwide. The majority of lung cancer cases are of non-small cell lung cancer type. With the advent of targeted cancer therapy, it is imperative not only to properly diagnose but also sub-classify non-small cell lung cancer. In our study, we evaluated the utility of using Apple Create ML module to detect and sub-classify non-small cell carcinomas based on histopathological images. After module optimization, the program detected 100% of non-small cell lung cancer images and successfully subclassified the majority of the images. Trained modules, such as ours, can be utilized in diagnostic smartphone-based applications, augmenting diagnostic services in understaffed areas of the world.

1. INTRODUCTION

**Background and significance**

Until recently, machine learning was a domain of data scientists. In 2018 we saw the creation of simplified models that gave developers and "citizen scientists" the ability to create sophisticated AI models with minimum effort and knowledge. Microsoft introduced Custom Vision[1] and Google introduced Auto ML [2] to fill these roles.

During the 2018 Worldwide Developers Conference (WWDC), Apple introduced Create ML. This new machine learning tool enables developers to import their data and create computer vision and natural language AI models on the macOS.[3] Create ML is built on Swift, Apple's programing language. It enables creation of AI models using the Xcode Playgrounds UI by simply dragging the training and testing folders into a model builder. Due to transfer learning technology that starts with a pre-trained model and uses the pre-trained model as a feature extractor, Create ML does not require as many images for training and testing. It also runs fast even on less powerful computers.

Models can be created in Create ML and run in Core ML 2, the next generation of machine learning technology that supports extensive deep learning with more than 30 layers along with standard tree ensembles, support vector machines, and generalized linear modes. Create ML Vision can be used to create an image classifier model which can classify images.

**Objective**

The purpose of our study was to evaluate Create ML Vision technology to detect and subclassify non-small cell lung cancer based on histopathological images.

## 2. MATERIALS AND METHODS

### 2.1 Data preparation

750 images (250 benign lung tissue, 250 adenocarcinomas, and 250 squamous cell carcinomas) were obtained using a Leica Microscope MC190 HD Camera (Leica Microsystems, Wetzlar, Germany) connected to an Olympus BX41 microscope (Olympus Corporation of the Americas, Center Valley, PA, USA) and the Leica Acquire 9072 software for Apple computers. All the images were captured at a resolution of 1024 x 768 pixels using a 60x dry objective and saved on an Apple MacBook Pro computer running macOS v10.13 (Apple Inc, Cupertino, CA, USA).

### 2.2 Xcode Image Classifier builder in Playground

We used the Apple Xcode Playground v10 beta on Apple MacBook Pro running macOS v10.14 beta to create ImageClassifier Model with following lines of code.

```
import CreateMLUI

let builder = MLImageClassifierBuilder()

builder.showInLiveView()
```

We opened the assistant editor in Xcode and then ran the code. The live view displayed the image classifier UI. We then dragged the training folder for training the model and the testing folder to evaluate the model on the indicated locations in live view. (Figure 1)

### 2.3 Experiment 1

In this experiment, we tested the Apple Create ML model to detect and subclassify non-small cell lung cancer histopathological images. We created three classes of images (250 images each) with the following labels: Benign (benign lung tissue), AdenoCA (adenocarcinoma), and SqCA (squamous cell carcinoma). The training folder included three class labeled subfolders with training images (80% of total). The testing folder included three class labeled subfolders with testing images (20% of total). To find optimal parameters for our model, we first tested it with the default parameters of 10 iterations and no image augmentations.

Next, we tested the model with different numbers of iterations.

Finally, we tested the model with the addition of various augmented images.

The best results were obtained with 50 iterations and no additional augmented images. We then trained and tested our model with 50 iterations, 10 times (Table 1).

| | Accuracy | (%) | | Recall | (%) | | Precision | (%) | |
|---|---|---|---|---|---|---|---|---|---|
| Run | Training | Validation | Testing | AdenoCA | SqCA | Benign | AdenoCA | SqCA | Benign |
| 1 | 100 | 97 | 95 | 94 | 92 | 100 | 92 | 94 | 100 |
| 2 | 100 | 97 | 96 | 94 | 94 | 100 | 94 | 94 | 100 |
| 3 | 100 | 97 | 96 | 94 | 94 | 100 | 94 | 94 | 100 |
| 4 | 100 | 97 | 96 | 94 | 94 | 100 | 94 | 94 | 100 |
| 5 | 100 | 97 | 95 | 94 | 92 | 100 | 92 | 94 | 100 |
| 6 | 100 | 94 | 96 | 94 | 94 | 100 | 94 | 94 | 100 |
| 7 | 100 | 97 | 96 | 94 | 94 | 100 | 94 | 94 | 100 |
| 8 | 100 | 94 | 97 | 94 | 96 | 100 | 96 | 94 | 100 |
| 9 | 100 | 94 | 97 | 94 | 96 | 100 | 96 | 94 | 100 |
| 10 | 100 | 97 | 95 | 94 | 92 | 100 | 92 | 94 | 100 |
| Mean | 100 | 96.1 | 95.9 | 94 | 93.8 | 100 | 93.8 | 94 | 100 |
| Median | 100 | 97 | 96 | 94 | 94 | 100 | 94 | 94 | 100 |
| Range | 100-100 | 94-97 | 95-97 | 94-94 | 92-96 | 100-100 | 92-96 | 94-94 | 100-100 |

Table 1. Testing model using 50 iterations as default.

### 2.4 Experiment 2

In this experiment, we tested the Apple Create ML model to differentiate between normal lung tissue and squamous cell carcinoma histopathologic images. We created two classes of images (250 images each) with the following labels: Benign (benign lung tissue) and SqCA (squamous cell carcinoma). The training folder included two class labeled subfolders with training images (80% of total). The testing subfolder included two class labeled subfolders with testing images (20% of total). We tested our model with 50 iterations, 10 times.

### 2.5 Experiment 3

In this experiment, we tested the Apple Create ML model to differentiate between normal lung tissue and adenocarcinoma histopathologic images. We created two classes of images (250 images each) with the following labels: Benign (benign lung tissue) and AdenoCA (adenocarcinoma). The training folder included two class labeled subfolders with training images (80% of total). The testing subfolder included two class labeled subfolders with testing images (20% of total). We tested our model with 50 iterations, 10 times.

### 2.6 Experiment 4

In this experiment, we tested the Apple Create ML model to differentiate between adenocarcinoma and squamous cell carcinoma histopathologic images. We created two classes of images (250 images each) with the following labels: AdenoCA (adenocarcinoma) and SqCA (squamous cell carcinoma). The training folder included two class labeled subfolders with training images (80% of total). The testing subfolder included two class labeled subfolders with testing images (20% of total). We tested our model with 50 iterations, 10 times (Table 2).

|     | Accuracy (%) |            |         | Recall (%) |      | Precision (%) |      |
| --- | ------------ | ---------- | ------- | ---------- | ---- | ------------- | ---- |
| Run | Training     | Validation | Testing | AdenoCA    | SqCA | AdenoCA       | SqCA |
| 1   | 100          | 95         | 93      | 94         | 92   | 92            | 94   |
| 2   | 100          | 95         | 93      | 94         | 92   | 92            | 94   |
| 3   | 100          | 95         | 93      | 94         | 92   | 92            | 94   |
| 4   | 100          | 95         | 93      | 94         | 92   | 92            | 94   |
| 5   | 100          | 95         | 93      | 94         | 92   | 92            | 94   |
| 6   | 100          | 95         | 93      | 94         | 92   | 92            | 94   |
| 7   | 100          | 95         | 93      | 94         | 92   | 92            | 94   |
| 8   | 100          | 95         | 93      | 94         | 92   | 92            | 94   |
| 9   | 100          | 95         | 93      | 94         | 92   | 92            | 94   |
| 10  | 100          | 95         | 93      | 94         | 92   | 92            | 94   |

Table 2. Lung adenocarcinoma vs lung squamous cell carcinoma

### 2.7 Experiment 5

In this experiment, we tested the Apple Create ML model to differentiate between normal lung tissue and non-small cell lung cancer histopathologic images with 50/50 mixture of adenocarcinoma and squamous cell carcinoma. We created two classes of images (500 images each) with the following labels: Benign (benign lung tissue) and NSCLC (non-small cell lung cancer). An additional 250 benign lung tissue images were created by horizontally flipping the original images. The training folder included two class labeled subfolders with training images (80% of total). The testing subfolder included two class labeled subfolders with testing images (20% of total). We tested our model with 50 iterations, 10 times.

### 3. RESULTS

When testing benign lung tissue against squamous cell carcinoma (experiment 2), adenocarcinoma (experiment 3), and against the combination of two non-small cell carcinoma subtypes (experiment 5), we obtained 100% recall and 100% precision for all classes of images. In other words, our AI model could correctly call benign lung tissue and malignant lung tissue of two types 100% of the time.

When testing the squamous cell carcinoma subtype of non-small cell lung cancer against the adenocarcinoma subtype of non-small cell lung cancer (experiment 4) we obtained a consistent overall 100% training accuracy, 95% validation accuracy, and 93% testing accuracy. The trained module classified adenocarcinoma images with a 94% recall and 92% precision, and squamous cell carcinoma images with 92% recall and 94% precision.

The most interesting and favorable results were obtained in experiment 1, in which we trained the model to detect non-small cell lung cancer and subclassify it into two subtypes: adenocarcinoma and squamous cell carcinoma. After optimizing the model, we obtained a median training accuracy of 100 %, median validation accuracy of 97%, and median overall testing accuracy of 96% (table 1). Our model was able to differentiate between benign lung tissue and malignant lung tissue 100% of the time. None of the benign images were diagnosed as cancer and vice versa. Furthermore, our trained model successfully subclassified the majority of non-small cell lung cancer with a median recall and precision of 94% for both adenocarcinomas and squamous cell carcinomas.

4. DISCUSSION

The onset and development of machine learning programs are rapidly changing many aspects of health care. Numerous studies involving artificial intelligence have been performed in the areas of dermatology [4-8] ophthalmology,[9] radiology[10-17] and pathology.[18-23] AI has also been utilized in the classification and detection of infectious diseases[24-25] as well as cardiology programs that assist in identifying patients with heart failure, improving cardiovascular risk predictions and improving heart failure survival analysis.[26-28]

The performance advantages of AI programs will be essential in worldwide modern healthcare.[29] Recent AI studies, attempted with various smartphone applications, have experimented with microscopy and diagnosis of dermatology lesions.[30]

Early detection, diagnosis, treatment options and prognosis of different cancers have motivated numerous AI healthcare studies.[19,23] Data analysis compiled from diagnostic images, genetic expression testing, and electrophysiological procedures, is transformed into valuable assets which may be utilized in treatment decisions, thus reducing errors and improving overall outcomes.[12]

It has been shown in previous dermatology studies, that AI programs demonstrated successful learning attempts in screening large numbers of data sets, providing further classification into subcategories that are then more easily diagnosed and interpreted.[4] Performed on large data sets, these pioneer studies provided further inspiration for useful deep learning studies on even smaller sets of dermatological clinical images.[5]

Lung cancer continues to be a major healthcare challenge. It is the leading cause of cancer death among men and the second leading cause of cancer death among women worldwide.[31] Non-small cell lung cancer represents 85 % of all lung cancer cases. Due to the recent availability of advanced targeted therapies, it is imperative to not only detect, but also properly subclassify non-small lung cancer into two major subtypes: squamous cell carcinoma and adenocarcinoma, which can be challenging even for experienced pathologists.[32-33]

In this study, we demonstrated the ability of an AI program to detect non-small cell lung carcinoma and successfully subclassify it.  Our study reveals the ability of an AI program to screen and diagnose non-small cell lung cancer with high percentages of sensitivity and specificity on small data sets of training images. Our trained module can be used in various AI iPhone applications. Developing technological advances, such as these, will become increasingly important and vital in remote areas of the world lacking accessible healthcare facilities. Diagnostic applications could even be run on low-cost, 3D-printed smartphone microscopes,[34] to help with the reported severe shortage of diagnostic pathologists in sub-Saharan Africa.[35]

5. **CONCLUSIONS**

Implementation of the machine learning algorithms may provide better detection, treatment, and prognosis of non-small cell lung cancer. Thus, greatly improving a patient's prognosis and survival rate.

**ACKNOWLEDGEMENTS**

None

**FUNDING**

This material is the result of work supported with resources and the use of facilities at the James A. Haley VA Hospital.

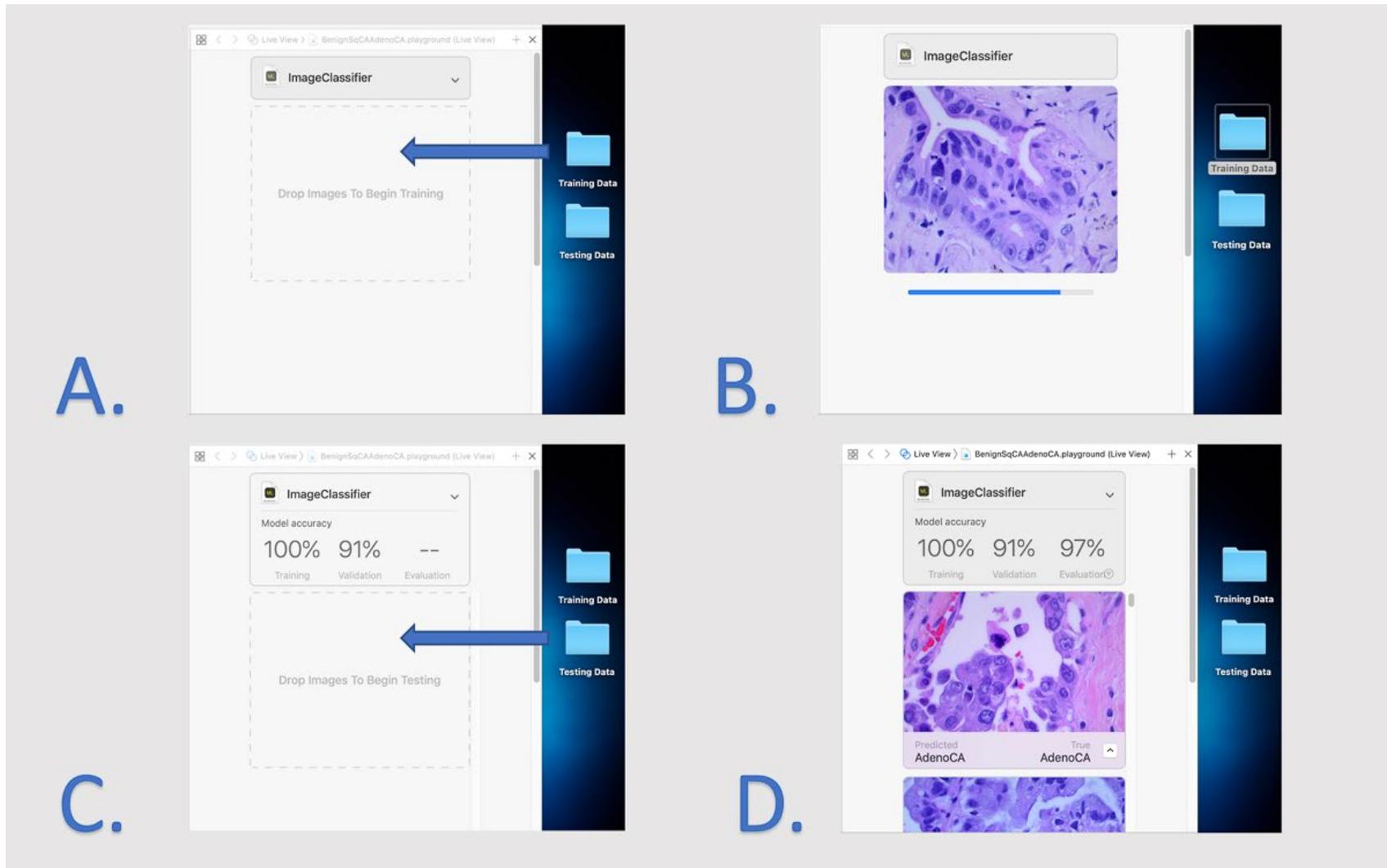

**Figure 1**. Apple Xcode image classifier user interface. Drag Training Data folder for training (A), training the model (B), drag Testing Data folder for evaluation of unknown images (C), end results for trained model (D).

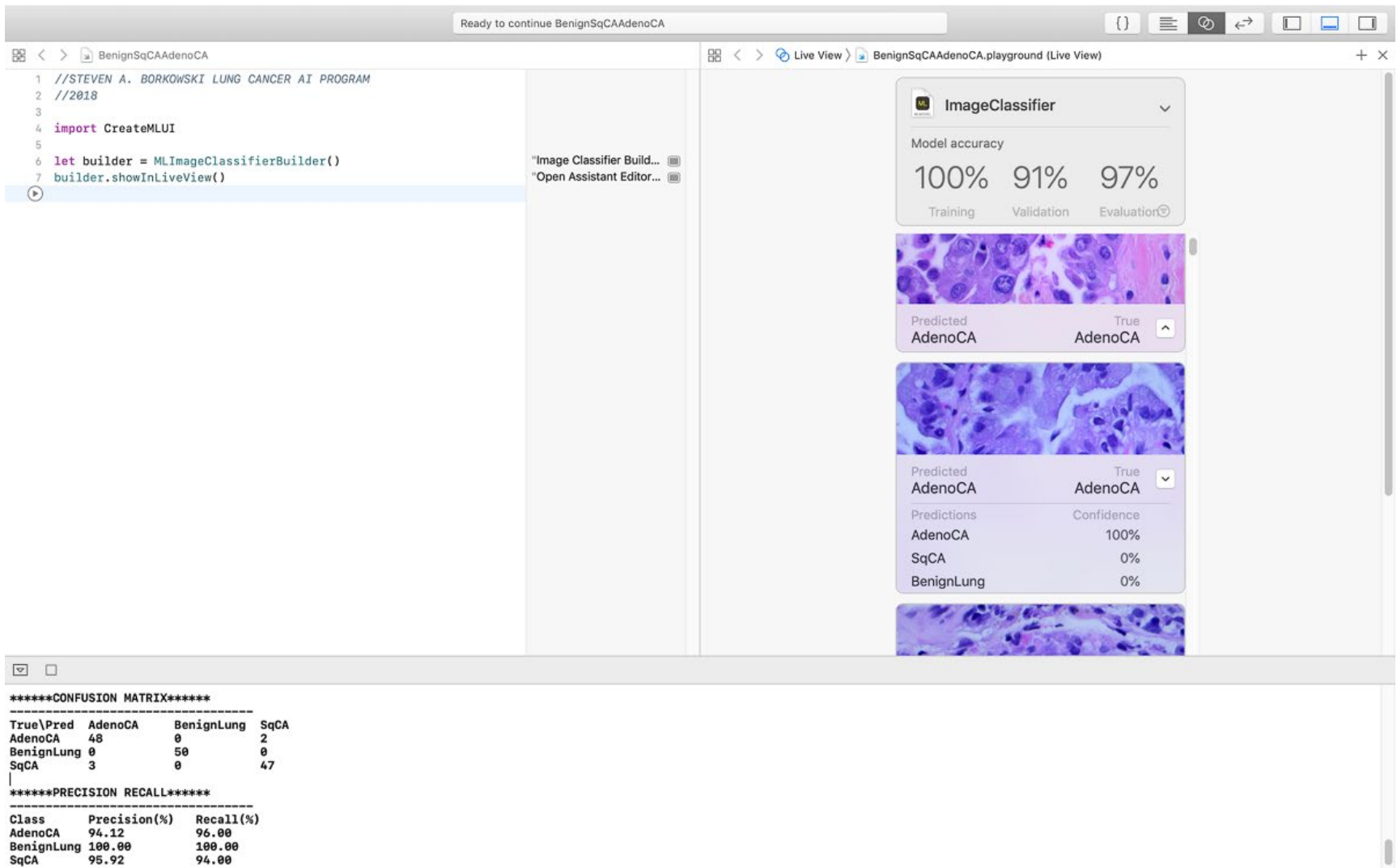

**Figure 2**. Trained model with console displaying confusion matrix and precision recall data.

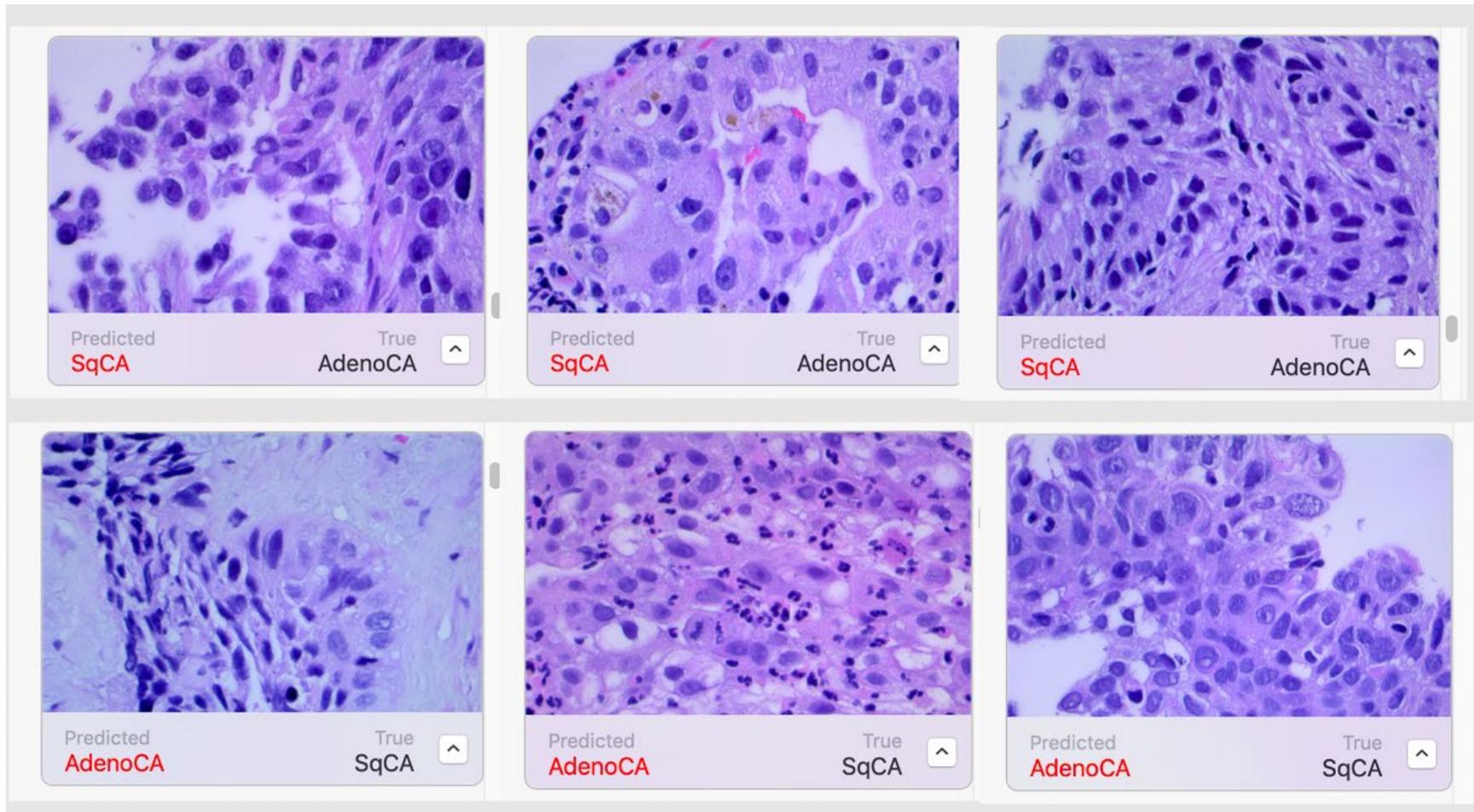

**Figure 3.** Examples of mistakes made by the model